# The demographics of physics education research

Stephen Kanim
Department of Physics, New Mexico State University, Las Cruces NM 88005
Ximena C. Cid
Department of Physics, California State University Dominguez Hills, Carson CA 90747

Abstract

Is physics education research based on a representative sample of students? To answer this question we skimmed physics education research papers from three journals for the years 1970 – 2015 looking for the number of research subjects, the course the subjects were enrolled in, and the institution where the research was conducted. We then compared the demographics of our research population to those of all students taking physics in the United States. Our results suggest that research subjects as a whole are better prepared mathematically and are less diverse than the overall physics student population.

Introduction

The purpose of this paper is to highlight, and to attempt to quantify, the disparities that exist between the level of preparation and background of the general population of American students taking introductory physics and the student population reported on in the physics education research literature. In general, physics education research (PER) subjects are better prepared mathematically than the overall population of introductory physics students. The selection of research subjects has primarily come about as a result of convenience: Since the bulk of the research has been conducted at more selective universities, and within these universities in more selective courses (*i.e.*, the calculus-based introductory sequence rather than from algebra-based, conceptual, or other courses), the level of preparation of most research subjects has been higher than the level of preparation of most students enrolled in physics courses. Effectively, the physics education research community has inadvertently cherry-picked its data.

The effect of a research focus on a non-representative research population is hard to assess. It might be that for some studies there is no effect at all, and repeating the same research on a more representative population would yield the same results. For example, we suspect that student responses to conceptual questions about Newton's 3$^{rd}$ law or about current in circuits would be similar across introductory student populations. On the other hand, questions that depend on an understanding of the relationship between a quantity and its rate of change might be strongly dependent on mathematical proficiency, as might questions about expectations, epistemology, or affect. But these are just guesses. If the *intent* of research-based curriculum development or the creation of theoretical models is to benefit physics students *in general*, then the PER community needs to be more attentive to demographics. On the other hand, if the intent is to describe, modify, and model the behavior of a subset of this population, then our community needs to establish norms for describing these subsets, and to be more specific about limitations on the applicability of our findings.

An additional unintended consequence of the selection of convenient research subjects is that fewer underrepresented racially and/or ethnically diverse student populations are part of the research population than are present in the general population of introductory physics students. In addition, the PER research population comes from wealthier families than the general population. As a result, both the challenges and affordances of more diverse student populations are not well represented in the research.

As with many systemic problems, these disparities between the PER research population and the general population of physics students are not a result of ill intent, and our goal in highlighting them is not to assign blame or to criticize past researchers for their choices of research subjects. We believe that the homogeneity of research populations has generated less ambiguous research results than would otherwise have been possible. In turn, the curricular materials developed on the basis of this research have been demonstrably successful at improving instruction at many institutions that are leaders in research and in curricular innovation. The rapid growth of PER over the years of our study, and the widespread adoption of research-based materials is in large part a consequence of limiting the variability in our student research populations.

PER is now reasonably well established as a field, however, and it is our sense that the time is ripe to start to explore the effects of variations in student population. Perhaps the most important reason for doing so is to improve instruction for *all* physics students by better understanding the impediments and affordances that students of diverse backgrounds bring to physics learning. In addition, efforts to systematically compare results across populations will allow refinement of our understanding of what matters in terms of preparation for physics instruction, and of which descriptors of student background are useful and which are not. A deeper understanding of variations in population and preparation will allow the development of curricular materials that are better matched to specific populations and that are more likely to be successful for the general population of physics students. Our understanding of the process of adoption of materials will benefit from paying attention to the effects of population differences. Finally, theoretical models of student thinking will need to become more nuanced in order to account for population variations, and the quality of these models will improve as they are tested against more general student populations. Exploring variation in population is a tremendous opportunity for the PER community, and will yield enormous benefits for our students and for our understanding of physics learning and teaching.

*Description of this study*

In order to evaluate what student populations the PER community has studied, we selected a sample of PER papers that we believed would provide a sufficiently broad but still manageable and representative sample of research descriptions. We chose to look at the research published from 1970 to 2015 inclusive from three journals: *The American Journal of Physics (AJP), Physical Review: Physics Education Research (PRPER),* and *The Physics Teacher (TPT)*. Here we describe our procedure, and in subsequent sections we summarize our results and the limitations of this study.

One author (Steve Kanim) scanned each physics education paper, looking for four elements: (1) The type of research study; (2) the total number of students that data were obtained from; (3) the institutions where the research was conducted; and (4) the courses that the students were enrolled in. (The topic of study was also noted, but is not relevant to this population study.) We used a very broad definition of physics education research to select which papers to scan, and many of the scanned papers were subsequently eliminated as described below. Because this scanning was relatively quick and in many cases we had to make guesses, results are necessarily rough.

To avoid eliminating a large number of papers, we developed some rules for dealing with uncertainties in the data. For cases where the paper we were scanning reported an aggregate number of students for different courses or for different schools, we simply split the number of students evenly among the courses or schools. When the name of the school was not given, we assumed that the research was carried out at the authors' home institutions. (If data was reported from multiple institutions without naming those institutions, we eliminated that paper from our data set.) For some papers we estimated the number of students from whom data was collected based on the reported number of sections and average class size. For example, a paper might say there were two sections of algebra-based introductory mechanics, which on average have 150 students. We would assume there were 300 students total even though there were likely fewer or more students for that particular study.

From a spreadsheet compiled from this data, we then eliminated studies that did not involve student data from physics courses (for example, purely theoretical papers, papers about textbook approaches, papers that described curricular or diagnostic modifications with only summaries of student success rates, or papers where all data was collected in mathematics or astronomy classes). Some papers included results from graduate students and faculty, and these populations were not included in our analysis. We also eliminated studies that were summaries of other research or were metastudies. In addition, we did not include studies or parts of studies that were conducted outside of the US. Finally, in some cases we could not make reasonable guesses about the information we were looking for based on our scanning of the papers, and in these cases we also eliminated the study.

From a total of 1,031 scanned papers, we included 417 papers in our data set, as shown in the top part of *Table I*. By our count, the included papers describe research conducted on a population of 257,657361 total students. There are likely cases of double-counting included in this total, because many studies report results from multiple questions, and it is often impossible to tell whether the same student population was involved.

|  | *PRPER* | *AJP* | *TPT* | Total |
|---|---|---|---|---|
| Papers: Total | 342 | 372 | 317 | 1,031 |
| Papers: Included | 179 | 159 | 79 | 417 |

| Student Population | N from *PRPER* | N from *AJP* | N from *TPT* | N Total |
|---|---|---|---|---|
| K-9 | 10 | 193 | 0 | 203 |
| High School | 674 | 1,590 | 19,275 | 21,539 |
| Two-year College | 380 | 321 | 0 | 701 |
| Teacher Prep | 938 | 1,598 | 260 | 2,796 |
| Disadvantaged | 0 | 69 | 0 | 69 |
| Conceptual | 615 | 2,597 | 469 | 3,681 |
| Honors | 405 | 924 | 0 | 1,329 |
| Algebra-based | 8,341 | 23,718 | 4,607 | 36,666 |
| Calculus-based | 56,991 | 109,218 | 13,487 | 179,696 |
| Upper level | 4,600 | 3,108 | 3 | 7,711 |
| Other | 145 | 3,061 | 60 | 3,266 |
| **Total** | **73,099** | **146,397** | **38,161** | **257,657** |

*Table 1: Number of papers included in this study and number of total students in various populations of courses from included papers.*

As shown in *Table 1*, we binned the students in each study into 11 categories based on their level and the courses in which they were enrolled: Kindergarten through ninth grade; high school; students at two-year colleges (enrolled in any physics course); students enrolled in pre- or in-service courses for teachers; students in special courses for disadvantaged or underprepared students; students in conceptual physics courses, students in introductory honors courses; students in introductory algebra-based courses; students in introductory calculus-based courses; students in upper-level physics courses; and students in other courses (*e.g.* lab courses). Students in all but the first two categories are in post-secondary courses.

In some studies it was clear that students were enrolled in an introductory course, but it was less obvious whether this course was algebra-based or calculus-based. For other studies, research subjects were chosen from both of these courses and the results were not separated. Rather than eliminate these studies, in both of these cases we arbitrarily chose to assign half of the students to each course. We compare these results to data about the relative sizes of the student populations obtained from the American Institute of Physics [1].

*Results*

*Overview*

Six important results are suggested by our study and are described below: (1) PER in the US pays scant attention to high school students; (2) PER in the US all but neglects students in two-year colleges; (3) PER studies of introductory courses focus on students in the calculus-based course; (4) PER studies are conducted at institutions where math preparation of incoming students is relatively strong; (5) PER studies are conducted at institutions that have wealthier students and have a smaller fraction of underrepresented racially and/or ethnically diverse students than the overall college-bound student population; and (6) Sampling in upper-division physics courses is also highly homogenized. These results are described in more detail below.

*Result 1: PER in the US pays scant attention to high school physics*

During the 2012-2013 school year, 1.38 million students were enrolled in physics courses in the US in both public and private high schools [1]. This segment of the overall physics student population is also growing the fastest over the past 20 years compared to introductory courses in colleges and universities, as shown in *Figure 1a*. In comparison, there were about 0.5 million students enrolled in introductory physics courses in colleges and universities. From our study, only 76 of 417 total papers (24 *PRPER*, 21 *AJP*, and 31 *TPT)* reported data collected from a total of 21,539 students (674 *PRPER*, 1,590 *AJP*, 19,275 *TPT)* in high school classrooms. Whereas about 3 of every 4 American physics students are studying in high school classrooms, fewer than 1 in 10 of our research subjects are high school students. Had we started our survey in 1972 rather than in 1970 this disparity would be even more pronounced: Over 10,000 of the high school students – about half -- are reported on in a single survey published in *TPT* in 1971.

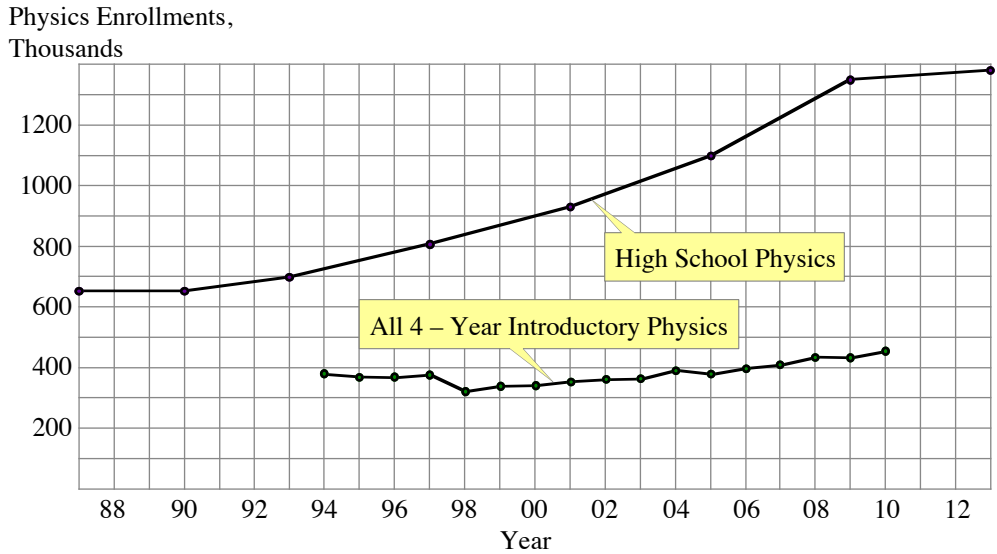

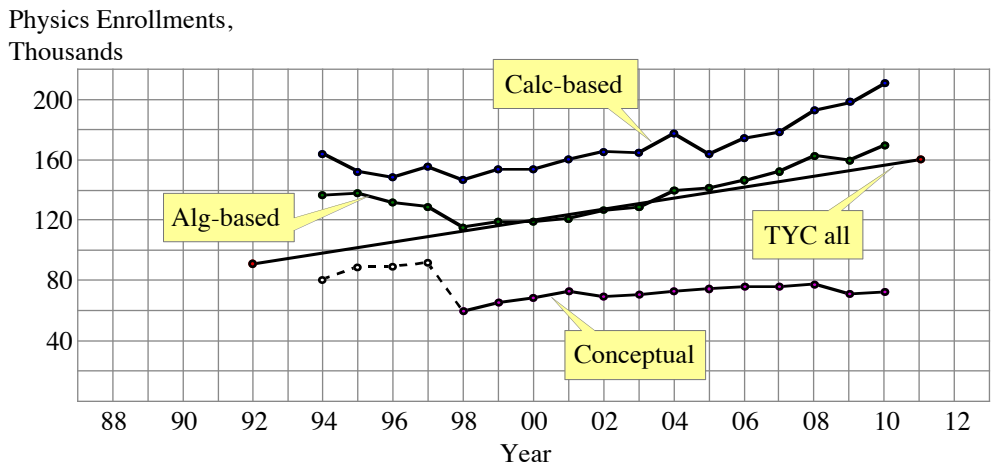

*Figure 1. (a) High school physics enrollment compared to physics enrollment at colleges and universities. (b) Enrollment in university courses broken down by type of course.*

This scarcity of data from US high school physics is distinctly different from PER conducted internationally, where it is much more common for research to be conducted in high schools. It is also different from mathematics education research in the US, which is primarily conducted in K-12 classrooms.

Of the six results that we report on in this paper, our claim about high school physics is most dependent on our choice of journals. Had we chosen *Journal of Research in Science Teaching* (*JRST*) instead of *TPT*, for example, we might have seen better representation from high schools in the research population. We discuss this limitation in more detail in a subsequent section of this paper.

*Result 2: In the US we do almost no research on physics students in two-year colleges*

About 44% of all undergraduates are enrolled in two-year colleges, and many of these students take physics. As shown in *Figure 1b*, in 2011 there were approximately 161,000 students taking physics at two-year colleges (calculus-based, algebra-based, and conceptual physics courses only), compared to about 450,000 students taking physics at 4-year institutions [1]. In our study we found only 6 papers (2 *PRPER*, 4 *AJP*, and 0 *TPT*) reporting on two-year college students in the United States, with a total student research population of 701 students (380 *PRPER*, 321 *AJP*, and 0 *TPT*). Although about one-quarter of all college physics students are enrolled in these institutions, only 0.3% of the total number of students studied by PER are from two-year colleges. Not only is data from two-year college students lacking in PER, but the American Institute of Physics has only collected data about physics in two-year colleges twice in the past 25 years. This scarcity of information severely detracts from our understanding of what happens in two-year colleges.

*Result 3: In introductory physics, PER tends to focus on students in the calculus-based course*

Our study also suggests a disproportionate reliance on data from the calculus-based course. As seen in *Figure 1b*, about 210,000 students were enrolled in calculus-based physics in 2010, accounting for about one-third of the students taking introductory physics of all kinds at colleges and universities. From our data set, there were 179,696 students enrolled in calculus-based physics, accounting for about 82% of all research subjects in introductory courses. Our results for introductory physics at the university level are broken down by course in *Figure 2*, and are compared to data about students enrolled in all courses.

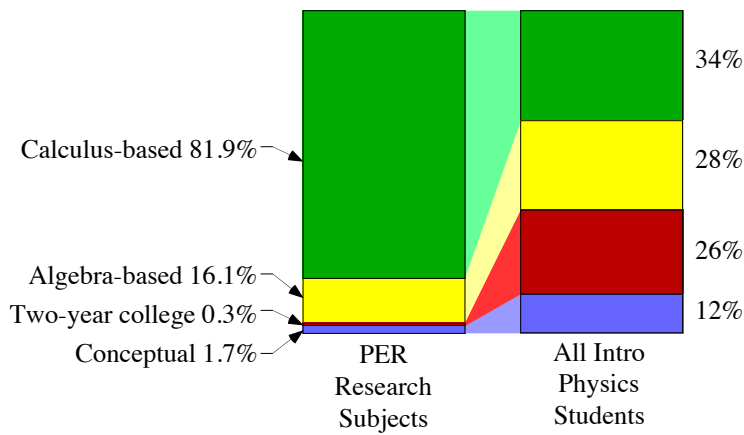

*Figure 2. Comparison of distribution of students in all introductory courses to distribution of all introductory students in our study. Here we have included students in honors introductory physics courses with students in calculus-based physics. About one-third of all students taking introductory physics in American universities take a calculus-based course at a 4-year institution. In contrast, more than four-fifths of all students reported on in the research papers of our study were in the calculus-based course at a 4-year institution. About one-quarter of all students taking introductory physics (of all varieties) do so at a two-year college; only 0.3% of the students in our study were from two-year colleges.*

*Result 4: In introductory courses, PER tends to focus on students with stronger mathematics preparation*

Next, we use our data to argue that PER has over-sampled students who are better prepared mathematically compared to the general student population of students taking the introductory course in college. This argument requires several assumptions that we enumerate as we describe our procedure below.

Because it is easy to obtain and relatively ubiquitous, we have chosen Math SAT scores [2] for entering freshmen for a university entering class as a proxy for mathematics preparation. Discussion of the relationship between SAT scores and physics performance are discussed in Refs. [3 – 5] (For all SAT Math scores discussed in this paper we have used scores from the 'old' SAT, administered before March 2016. For a few schools we were unable to obtain SAT scores and substituted converted ACT Math scores.) We acknowledge that, in practice, SAT Math scores are at best a crude measure of the degree to which students are prepared to use math *in physics* and look forward to the development of more relevant measures.

We expect that the SAT Math scores for students enrolled in introductory physics courses (algebra-based sequence and calculus-based sequence) at any given university are higher than those scores for the freshman class as a whole (except institutions where all freshmen are required to take introductory physics), and we are assuming that the difference is about the same for all universities. We offer a crude test of this assumption in the *Limitations* section of this paper.

From the 257,657 students included in this overall study, we eliminated the K-12 students and the students in 'other' courses, leaving 224,938 students in college courses of all types. We further eliminated students from unnamed institutions, leaving 210,784 students. (While entire studies without named universities had previously been eliminated from our study, these additional students were from studies where some of the data was from named institutions, and other data was not.) We then sorted the universities in our sample by the number of students that had been cumulatively reported on in their studies. The 39 universities with the largest number of students account for 95% of all the US university students reported on in the three journals, 200,079 of 210,784 students. For these 39 universities, we looked up the $25^{th}$ and $75^{th}$ percentiles of SAT math scores for the entering freshman class for the year 2016 [2].

In *Figure 3*, we illustrate disparities between the SAT scores of the student research population and the overall distribution of SAT scores for students in introductory physics. The blue histogram shows the SAT Math scores for all students taking the exam in 2008, with the middle half shaded a darker blue. The $25^{th}$ and $75^{th}$ percentiles for all students taking the SAT are indicated by the dashed vertical blue lines (430 and 590 respectively).

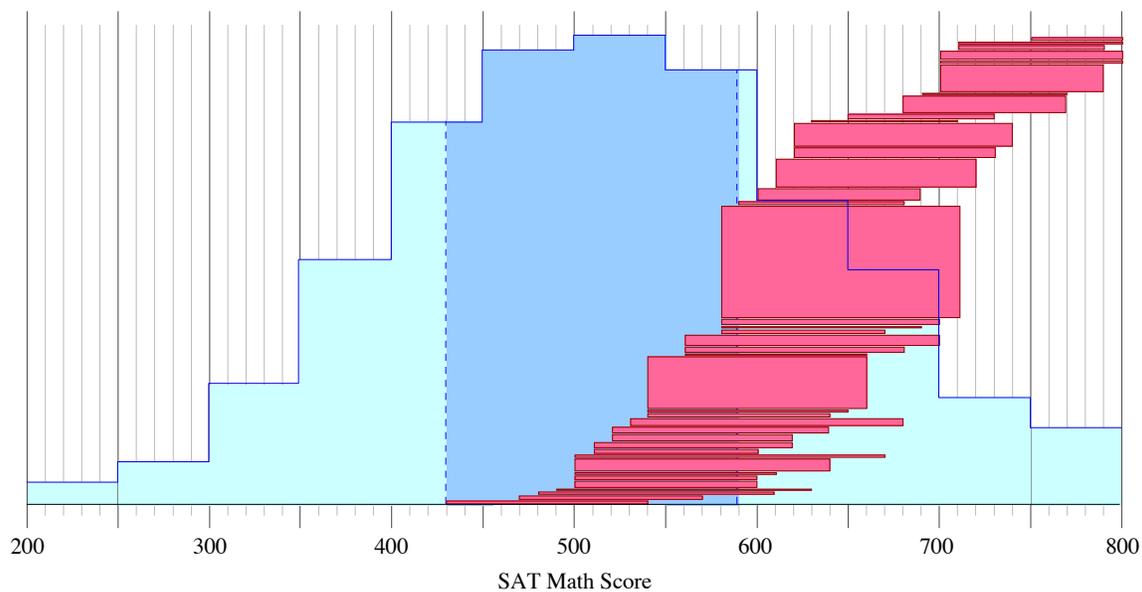

*Figure 3. Comparison of middle half SAT Math scores for all students with those of incoming freshmen at universities where most PER is conducted. The blue-shaded histogram gives the distribution of SAT math scores for all students taking the SAT in 2008 [6], the last year that the College Board released this data in 50-point bins. The dashed vertical blue lines represent the 25th percentile score (430) and the 75th percentile (590) scores for all students; with the middle half shaded a darker blue than the lowest and highest quarters. The 39 red-stacked rectangles are the 25th – 75th percentile ranges for incoming freshmen at the universities whose students comprise 95% of all university students in our study. The horizontal span of each rectangle in the stack represents the middle half of incoming freshman students for a single university, with the height of that rectangle giving the proportional representation from that university of all students in our study. The rectangles are stacked in order of increasing 25th percentile scores for each school to maximize overlap in the figure with the middle half of all students.*

The red-shaded area represents the physics education research student population from our data. Each of the 39 universities with the most students in the research population is represented by a rectangle extending horizontally from 25th percentile to 75th percentile for that university. The relative heights of the rectangles are determined by the fraction of the total research population represented in studies from each university.

A second comparison can be made from the binned SAT Math scores for each school's incoming freshmen [2]. For each of the 39 schools, we used College Board data and the total number of students in our study from that school to calculate how many of the students were in each 100-point bin. For example, if we had 1,000 students from a school, and 20% of the incoming freshmen from that school had SAT Math scores between 300 and 390, then we counted 200 students from that school with scores in this range. We then added up our binned data for all 39 schools to obtain the percentage of students in our study with scores in each bin. Since the College Board also reports binned data for all students taking the test [2], we can compare the results, as shown in *Figure 4*. Again we see that the PER sample is not representative: For example, while 46% of the overall population has SAT math scores below 500, only 7% percent of the students at PER research institutions do. At the other end of the scale, 25% of the overall population has SAT math scores of 600 or more, but students with these scores account for 70% of our research institutions.

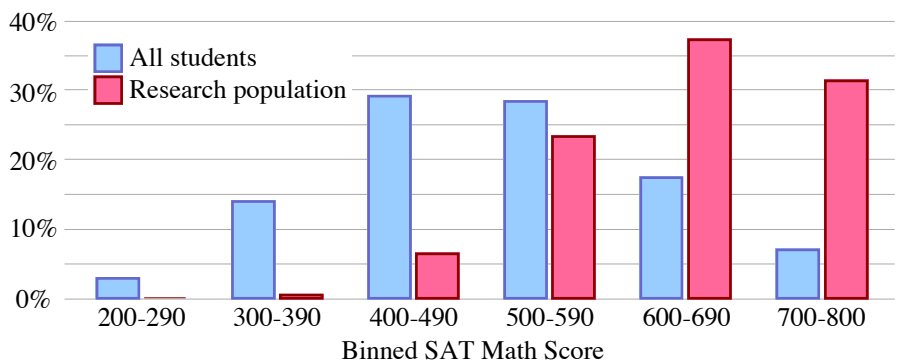

*Figure 4. Binned comparison of SAT Math scores for all college-bound students with students at universities where most PER is conducted. The blue bars are the SAT Math scores for all students who took the SAT in Spring 2016, collected into 100-point bins and given as a percent of all students taking the exam. The red bars are for students at the 39 research universities where most PER is conducted, weighted by the contribution of each university to the total number of students in our sample.*

We recognize that not all high-school seniors who take the SAT will enroll in college, and that not all students who enroll in college will take physics courses. Nonetheless, for many universities that are trying to improve their physics instruction, the 75th percentile for their students is about where the 25th percentile lies for many universities conducting physics education research. For example (and to bring this argument back to its genesis) the SAT Math distribution for entering freshmen at NMSU looks very much like the overall distribution for high school seniors. For instructors at these universities, it requires a substantial leap of faith to assume that typical PER research results have much relevance for their student populations.

*Result 5: PER in the US oversamples from white and wealthy populations*

PER researchers by and large conduct research on students at their own institutions, so the demographics of the research population reflect those of the institutions conducting research and are not necessarily similar to those of the overall population of physics students. *Figure 5* below was generated by looking at the racial demographics of the same 39 institutions that make up 95% of our student populations represented in the PER literature in our study. As with the SAT Math data, we scaled the demographics reported from each institution [2] by the fractional contribution of that institution to the overall PER student population. We then compared the racial demographics of the overall research population to the racial demographics of all college-bound seniors for 2015, also obtained from the college board [2]. The categories in *Figure 5* match the categories presented by the College Board website.

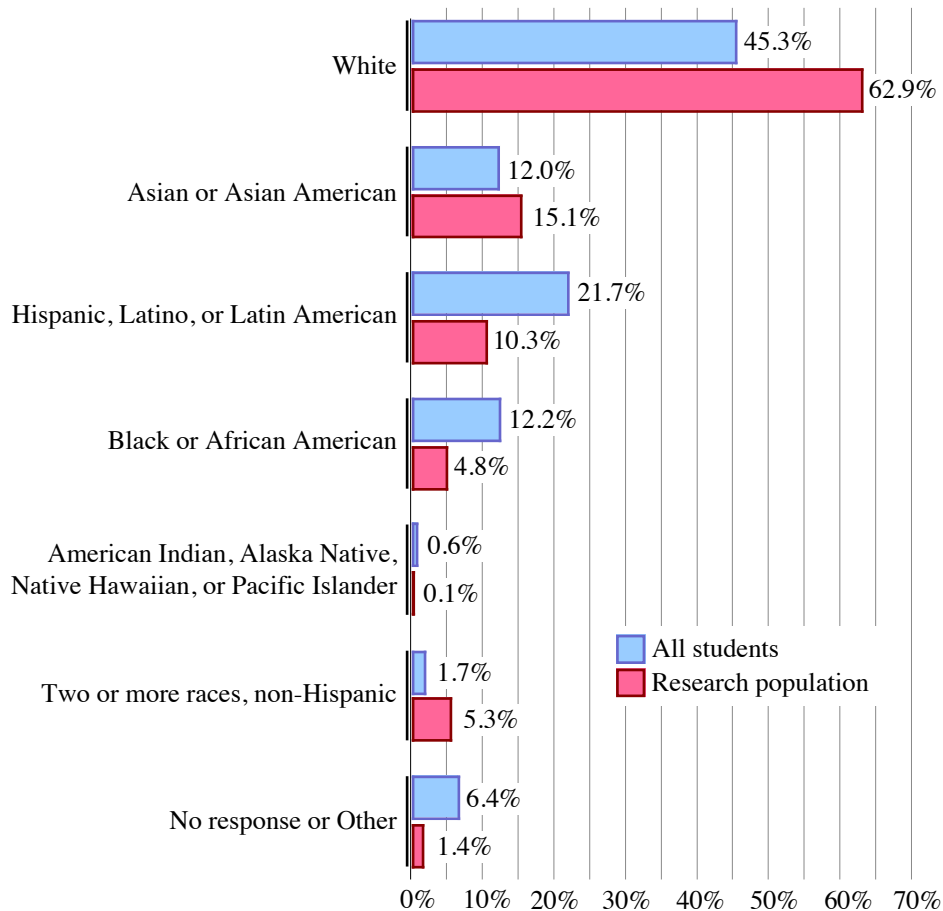

*Figure 5. Comparison of racial/ethnic distributions of all students taking the SAT in 2015 versus the PER research population from our study. Data and category titles are taken from the College Board website.*

The PER data is non-representative of underrepresented racially and/or ethnically diverse populations: While Latina/Latino, Black, and Indigenous American students are 34.5% of the college-bound students taking the SAT, by our estimate only 15.2% of the research student population are from these groups. The only exception is the category for *Two or more races, non-Hispanic*. We can only speculate about this unexpected result.

As with our discussion of math preparation, it is likely that the racial or ethnic makeup of all incoming freshmen is different from that of students in physics courses. Based on the demographics of STEM fields in general, it is probably true that there are fewer underrepresented students in physics courses than there are in the overall student population at any university. We don't know whether the difference is greater for the 39 schools in our sample than in all universities.

While there are PER groups at Hispanic serving institutions (HSI) and minority serving institutions (MSI), the data reported from these institutions is less than 5% of the overall data collected for physics education research as represented by our 39-school sample. Even for research from these institutions, the sample demographics are often either not included or are included as a description of the student population without discussing how demographics influence results.

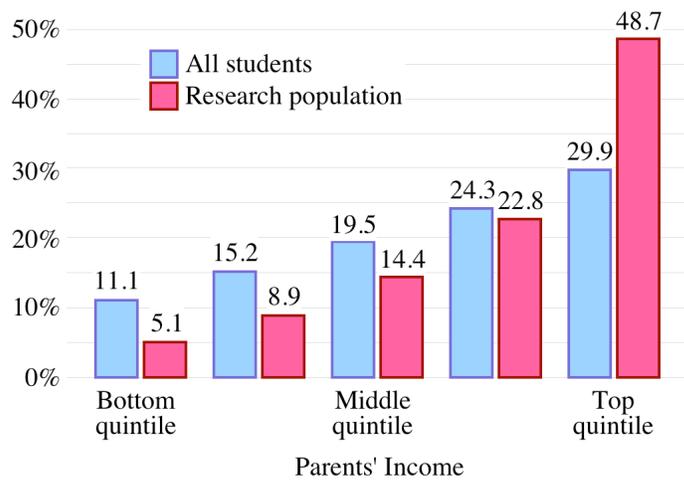

*Figure 6. Comparison of parents' income for research student population to overall student population.*

A similar analysis can be made comparing the research student population with the overall student population broken down by parents' income as shown in *Figure 6*. Data from the Equality of Opportunity Project [7] divides parents' income for students at each school into quintiles -- for example, giving the percent of students at a school whose parents' incomes are in the bottom quintile of all parents' incomes in the country.

Data was not available for two of the 39 schools in our research population, representing a total of about 8% of the research student population. We took the data for the other 37 schools, representing 87% of the total research student population, weighted the percentages by the fraction of students from that school in the overall research population, and added the results to obtain a parent income distribution for our research population.

From the list of all schools included in the Equality of Opportunity Project study, we eliminated categories of schools that we though were unlikely to offer physics – schools categorized as 2-year for profit, and schools that offered programs of study lasting less than two years. We compared the results from our research population to the percentages in each quintile of the remaining schools included in the study weighted by student population. (Repeating the comparison while including all schools made almost no difference in our results, with the largest changes to any quintile of about 0.3%.) Results are shown in *Figure 6*.

For eight institutions on our list, representing about 23% of the total research population, data was given for a university system rather than for a specific university *(e.g.,* cumulative data is presented for all branches of state university system rather than for individual universities within that system). In each of these eight cases, the research data was collected at the flagship university of the university system, and it is likely that research students are wealthier than the data suggests.

Almost half of the students in our research population have parents with incomes in the top quintile of all parent incomes, compared to about 30% of the overall population. At the lower end of the income scale, 14% of the students in the research population have parents with incomes in the bottom 40%, whereas about 26% of the students in the overall population fall into these categories. Overall, students in our research population come from wealthier families, and probably were educated in wealthier school districts.

*Result 6: Sampling in upper-division physics courses is highly homogenized*

Our data includes results from 6,722 students in upper-level physics courses. (Again, this is after removing students from unnamed institutions.) Only 22 different universities are represented, and over 95% of these students attend 9 universities. Since the two universities reporting the most students account for 80% of the upper-level students included in this study, the demographics of these two universities strongly determines the demographics of physics education research conducted in upper-level courses, and by extension for topics such as quantum mechanics and statistical mechanics.

Again using a proportional distribution of expected SAT Math scores as we did in our description of Result 3, we find that about 58% of the research sample of upper-level physics students can be expected to have scores above 600, compared to about 25% expected of students at all schools. This result is particularly salient for upper-level students in light of a study [4] that claims that a practical minimum SAT Math score of 600 is necessary for students to succeed as physics majors.

We expect that about 60% of our sample population of upper-level students will be White (compared to 45% of all students), about 11% Black or Latinx (compared to 34%), and a reasonable number of Native Americans to expect in our upper-level research sample is zero.

*Limitations of this study*

In this section we describe what we believe are the most important sources of uncertainty and error in our study, and what we did to gauge the impact of these uncertainties.

*Limitation 1: Errors and uncertainties introduced by our data collection methods*

This study was initiated by author Steve Kanim, who scanned through all papers from PRPER, AJP, and TPT, with results shown in *Table 1*. As described earlier, only enough time was spent on each paper to find: (a) the type of research study, (b) the total number of students involved in the study, (c) the institutions where the research was conducted, and (d) the course that the students were enrolled in. To evaluate the correctness of these numbers, the authors chose a random subset of 300 papers (100 from PRPER, 100 from AJP, and 100 from TPT), and Ximena Cid independently scanned these papers so that we could compare results.

In our individual scans of the papers, it was unusual for either author to spend more than 5 minutes on any given paper. We recognized at the outset that, because we were not reading the papers in detail, we could expect to make errors in our data tabulation. We decided in advance that when a disagreement occurred we would read the paper in more detail, and then arrive at a consensus. Often the information we sought was not explicitly provided as part of the paper, and we had to make judgments or guesses. For these reasons, we expect that any attempt to replicate our results would create a different data set. Our hope is that the conclusions that we draw are sufficiently robust that a replication would not substantially change the conclusions drawn.

Often we disagreed because we looked in different places for the information we sought. For example, data on the number of students in a study appears in a variety of places: abstract (sometimes an approximation), in introductions, in descriptions of methodology, in tables, in table captions, in figures, in figure captions, in the body of the text (as numbers, words, or both). It is very common for a single paper to include multiple studies and multiple reports of data, and perhaps the most common source of disagreement and ambiguity was determination of whether data from students in one study were from the same students as in another study reported in the same paper. Additionally, it was also often unclear if pre- and post-data was from the same sample population, and pre- and post-data often report different numbers of students, and for some studies they appear to be pre-treatment and post-treatment values for completely different populations altogether.

For our consensus data we settled on the following rules in addition to the rules enumerated in the section of this paper describing the study above: For cases where pre and post data reported different $N$ values, we choose the larger of the two reported numbers. When the name of the school was not given, we assumed that the research was carried out at the author's home institution. If data was reported from multiple institutions without naming those institutions, we eliminated that paper from our data set. For some papers we estimated the number of students from whom data was collected based on the reported number of sections and average class size (for example, a study might say: data was taken from 3 lecture sections that have on average 200 students per section).

The 300 papers we both reviewed represent a sampling of 30% of the 1031 total papers that form our data set. Almost all of the differences in our tabulations were in the number of students, rather than in the course category or in the institution represented. Often, we made different decisions about whether multiple sets of reported data represented the same group of students.

*Table 2* represents the consensus total for the number of students in each category after more detailed reading of the papers. The table entries with white backgrounds are those that were unchanged from the values reported in *Table 1*; the entries with colored backgrounds have the percent change indicated. Some of these changes for a single journal are quite substantial – in the worst case, a 32% increase (though this is for a small $N$ value for the "other" category). On the other hand, the total number of students for any given student population for all journals combined was 18% or less in all cases. Our sense is that the conclusions we have drawn about population disparities are not that sensitive to the variations in categorization and counting due to differing interpretations of paper descriptions, or due to overly hasty scanning of papers.

| Student Population | N from *PRPER* | N from *AJP* | N from *TPT* | N Total |
|---:|---|---|---|---|
| K-9 | 10 | 193 | 0 | 203 |
| High School | 674 | 1,590 | 19,292 | 21,556 |
| Two-year College | 380 | 321 | 0 | 701 |
| Teacher Prep | 1,188 (+27%) | 1,623 (+2%) | 260 | 3,071 (+10%) |
| Disadvantaged | 0 | 69 | 0 | 69 |
| Conceptual | 803 (+31%) | 2,692 (+4%) | 519 (+11%) | 4,014 (+9%) |
| Honors | 523 (+29%) | 816 (−12%) | 0 | 1,339 |
| Algebra-based | 8,505 (+2%) | 24,139 (+2%) | 4,609 | 37,253 (+2%) |
| Calculus-based | 65,308 (+15%) | 108,907 | 13,426 | 187,641 (+4%) |
| Upper level | 5,594 (+22%) | 3,489 (+12%) | 3 | 9,086 (+18%) |
| Other | 191 (+32%) | 3,356 (+8%) | 82 (+37%) | 3,629 (+11%) |
| **Total** | **83,176** (+14%) | **148,274** (+1%) | **38,191** | **269,641** (+5%) |

Legend: 1% – 5%, 5% – 10%, 10% – 20%, > 20%

*Table 2: Data corrections based on review of one-third of the papers. Uncolored entries remained unchanged by review, or were changed by less than one percent. While changes to individual entries from single journals were sometimes changed by up to 32 percent, total changes to individual categories had a maximum change of 18 percent.*

*Limitation 2: Effect of limiting our study to three journals*

The claim we make in this paper is that the research basis for PER in the United States is not based on a representative sampling of the students enrolled in physics. However, this study looks at PER only in the *American Journal of Physics, Physical Review – Physics Education Research,* and *The Physics Teacher*. How well do these three journals represent the publication record of all of PER? Is it possible that a wider sampling of journals would change our claim? To attempt to answer these questions, we looked at two summaries of PER with extensive publication listings: Most recently, a paper by Docktor and Mestre [8] and an earlier resource letter by McDermott and Redish [9].

*Docktor and Mestre synthesis*

In a 2014 paper, Docktor and Mestre [8] use the results of a paper commissioned by the National Research Council to summarize and categorize physics education research. The paper has 539 references, of which 84 are books, 40 are conference proceedings, 26 are websites, 10 are dissertations, and 2 are videos. The remaining 378 references are journal articles from a total of 41 journals. Of these, 250, or about two-thirds, are included in our 3-journal sampling. Twenty of the remaining 38 journals have only one or two papers listed from those journals. Four journals are reasonably well represented: the *Journal of Research in Science Teaching, JRST,* (20 papers), *Cognition and Instruction* (12 papers), the *Journal of the Learning Sciences* (12 papers, one of which is in reference to an entire issue), and *Science Education* (11 papers), collectively accounting for 55 of the 128 papers in other journals. For comparison, the Docktor and Mestre summary cites 133 papers from the *American Journal of Physics*, 100 papers from *Physical Review Special Topics – Physics Education Research*, and 17 papers from *The Physics Teacher*.

*McDermott and Redish resource letter*

In 1999, McDermott and Redish published a physics education resource letter [9] containing 243 references. (Some footnote numbers reference more than one paper.) Of these, 52 are books, 10 are conference proceedings, and 3 are websites, leaving 178 references to journal publications distributed over 25 journals. Not quite half (86 or 48%) of these publications are included in our 3-journal study, 64 papers in the *American Journal of Physics* and 22 papers for *The Physics Teacher*. (Because this resource letter was published in 1999, there are no reference papers for *PRPER*, which began publication in 2005.) Fifteen of the remaining 22 journals have only one or two referenced papers. The four most commonly cited of the remaining journals are the *International Journal of Science Education* (32 papers), the *Journal of Research in Science Teaching* (16 papers), *Physics Education* (9 papers), and *Physics Today* (7 papers).

Based on these two PER summaries, our 3-journal choice is reasonably effective at representing PER publications during the years 1970-2015. It might have been better to have also included the *Journal of Research in Science Teaching*, as papers there were cited 38 times in the two summaries, about the same as the 39 citations for papers from *The Physics Teacher*. We can get some sense of the effect of this inclusion by looking at the 38 citations from *JRST*. Only two papers were cited in both summaries, so there were 36 distinct papers included. Of these, 5 papers were summaries or theory papers that included no student data, 11 papers had no data from students in the United States, one paper included data only from students not taking physics, and one paper was based on data from graduate students. The 18 remaining papers were based on 4,186 research subjects.

Fifty-eight percent (2,437 subjects) were elementary- and middle-school students, reported on in two studies about scientific reasoning in the late 1970s. In contrast, there were only 203 K-9 students among the subjects of the papers in the 3 journals we chose. Almost 20% (815 subjects) were high school students. This more than doubles the percentage of high school students who were research subjects in our study of *AJP*, *TPT*, and *PRPER* papers. It is safe to say, then, that inclusion of *JRST* would have weakened our claim that PER in the United States pays scant attention to high school students. If we assume (based on the two summaries described in this section) that the inclusion of *JRST* would add about the same number of papers and therefore about the same number of research subjects to the pool as *TPT* does, and that the number of research subjects from high schools in *JRST* remains 20%, then we can expect that the total fraction of high school students in an expanded study would be about 10%, rather than 8%. Keeping in mind that about three-fourths of all introductory physics students are in high school, our conclusion (Result 1) that PER is dramatically under sampling high school students is likely unchanged by inclusion of *JRST*.

Only 899 of the *JRST* subjects, or 21%, were university students. All but 57 of these students were from the 39 schools described previously that make up the bulk of PER research subjects. There were no studies that included upper-level students. We do not believe that including *JRST* would have resulted in any changes to our five other research results, because the students by and large come from the same schools and therefore have similar demographics as the students that we are basing our conclusions on.

With more time, a valuable follow-on study would be to look at the effects of including *JRST* and perhaps other journals in more detail. Based on our analysis above of a small sample of these additional studies, though, we do not expect that expanding the study in this way would have much effect on the results we describe in this paper. An interesting additional study would be of the degree to which the PER research community overlaps the high school science learning research community, perhaps as measured by intra- and inter- citation instances for these two groups.

We note in closing that, while it might seem beneficial to add journals to our study, there is also an inherent confounding factor: Many of the papers that were included in the two syntheses but published in other journals were based on the same research studies as those already published in the 3-journal analysis that forms the core of our discussion, with the authors providing an alternate analysis, or perhaps one better suited to a different audience. As a result, a number of the students included by adding journals would be double-counted.

*Limitation 3: Use of university SAT Math data instead of physics class data*

In this paper we have used SAT Math scores obtained from the College Board as a measure of the mathematics preparation of students in physics courses at the 39 institutions where physics education is primarily conducted. This data is for all incoming freshman at each institution, and we assume that at almost all institutions the SAT Math scores are higher for students enrolled in physics courses than those scores for all freshmen. Result 4 (research students are better prepared than the overall population of physics students) and to a lesser extent Result 6 (research on upper level physics students is highly homogenized) depend on an assumption that SAT Math scores of physics students are monotonically related to those scores for incoming freshmen. With this assumption, (since the upward shift in scores happens for all schools) a comparison of incoming freshmen scores will inform us about the relative scores of students in physics courses.

This assumption may not be valid for schools with extremely high SAT Math admission scores. For a hypothetical school with a $25^{th}$ percentile of 700 and a $75^{th}$ percentile of 790, for example, we would not expect the scores in physics courses to be much different. (Some STEM-oriented schools require that all students take physics; for these schools the incoming freshmen would have the same score distribution as students in physics.)

For the university with the lowest $25^{th}$ and $75^{th}$ percentiles of the 39 schools that comprise our PER research sample (the bottom red rectangle in Figure 3) we would expect no such ceiling effect. This university uses ACT scores rather than SAT scores, and we have converted them to SAT equivalent scores in figure 3 and below. We looked at two fall semesters that included four sections of the algebra-based introductory physics course (N = 438) and four sections of the calculus-based introductory physics course (N = 507). We compared SAT Math scores for these classes to the SAT Math scores for incoming freshmen at that university during the same two semesters. Both the 25th and 75th percentile scores were about 30 points higher in the algebra-based course than for all incoming freshmen. In the calculus-based course the 25th percentile was about 90 points higher than the 25th percentile for all freshmen; the 75th percentile was about 70 points higher. Roughly, then, compared to all incoming freshmen the algebra-based middle half had SAT Math scores about 30 points higher and the calculus-based course had scores about 80 points higher. The median scores also increased by about the same amount.

We also had access to data from a calculus-based course (N = 277) at a university whose SAT Math scores placed about halfway up our list, the 22nd of 39 schools ranked by increasing 25th percentile scores. The 25th percentile for the SAT Math was 650 in the calculus-based course, or 70 points higher than for all freshmen at this university. The 75th percentile was 720, only about 20 points higher.

Using the data points for the calculus-based courses (about 82% of the research population), and assuming that the school with the highest SAT Math scores will have no appreciable difference between the scores of students in their physics courses and the scores of all students, we can make some very crude guesses about how Figure 3 might change if we had data for physics classes rather than for incoming freshmen. With 3 25th percentile score conversions (430 for all freshmen to 520 for physics students; 580 to 650; 750 to 750) we can fit a quadratic to our data, and use this quadratic to map the 25th percentile scores for all schools. We can do the same for the 75th percentile (540 to 610; 700 to 720; 800 to 800). We use the same mapping for the 25th and 75th percentile range for all physics students. The results of this mapping are shown in Figure 7.

The most noticeable result of our mapping is a narrowing of the score range for the middle half of students (both for the research population schools and to a lesser extent for the overall student population, since the 75th percentile increases less than the 25th percentile. There is relatively little change in the overlap between the research population and the overall student population, and there is still a clear mismatch in levels of preparation as measured by SAT Math scores. We do not believe that our conclusions about this mismatch would change if we had access to scores for physics students rather than for incoming freshmen.

As a final note about uncertainties in SAT Math scores, our use of incoming freshmen scores overlooks the scores of students who transfer to four-year institutions from two-year colleges. About one-quarter of the students who enter a two-year college transfer to a four-year institution within five years. Some of these students will have already taken physics when they transfer, and some will take physics after transferring. Depending on institutional admissions policies, these students may not be using SAT scores (even if they did take the SAT in high school) for their admission on transfer. Without more information, it is hard compare the level of preparation of these students to non-transfer students.

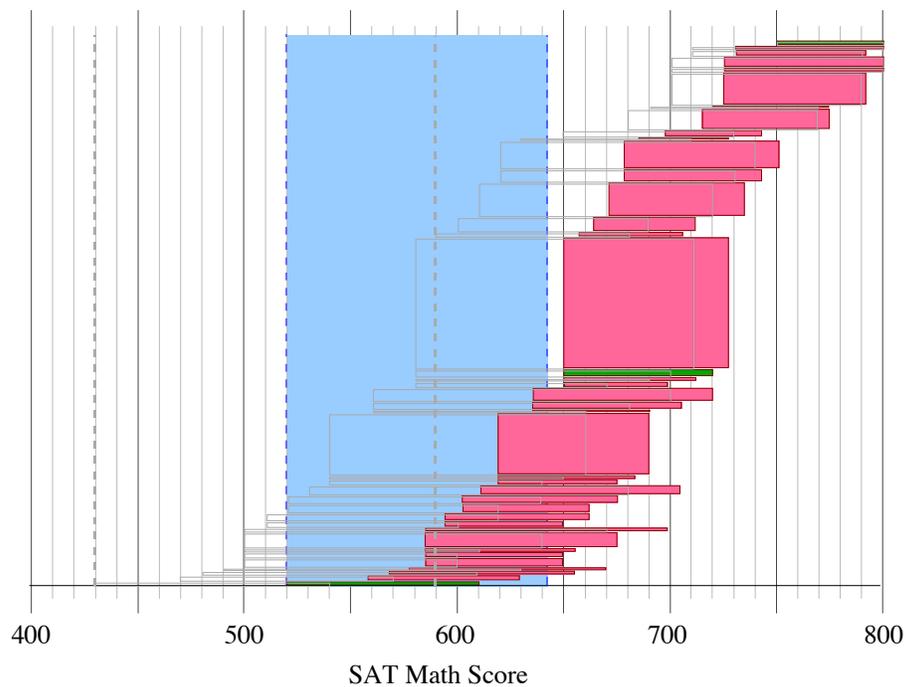

*Figure 7*. Our best guess about the overlap between research population and overall physics student population. Based on our knowledge of how SAT Math scores change from incoming freshmen to students in calculus-based physics courses for 2 institutions (shown in green), and on the assumption that the scores do not change for the school with the highest scores (also in green), the 25th to 75th percentile ranges shown in Figure 3 have been mapped to predicted ranges for students in calculus-based physics courses. Score mapping is done by quadratic fit. The data for incoming freshmen from Figure 3 is shown outlined in light grey for comparison, as are the limits for the middle half range.

The second issue is that we are not taking into account those students that are transferring from two-year colleges. Students from TYCs might have taken the introductory physics sequence before transferring or might transfer and take the introductory sequence at the four-year institution. Depending on institutional admissions policies, these students may not be using SAT scores (even if they did take the SAT in high school) for their admission on transfer, nor have these students been identified in the studies we have sampled. We believe that the transfer rates are small enough that any effects do not alter the conclusions we have made.

*Discussion*

As university students, some of us have had the experience of serving as research subjects for psychology and behavioral science studies: Often, students in general education psychology classes are asked to sign up for these studies as part of their course credit. For researchers, these students form a convenient subject pool, because they are easy to recruit and are already on campus. But do they form a representative sample? In a paper titled *The weirdest people in the world?*, [10] J. Henrich, S.J. Heine, and N. Norenzayan look at the research basis for behavioral science, which as a field almost exclusively chooses sample populations from Western, Educated, Industrialized, Rich, and Democratic (WEIRD) populations. The authors point to various cross-cultural studies that show that the responses obtained from western college students are not predictive of the responses of humans in general, and in fact are often outliers. This does not mean that the results of studies depending on these students are incorrect or that they are not useful. It does mean, however, that behavioral scientists are making claims that probably are not universally generalizable when they describe results so obtained as having general validity, or when they use the results of these studies to make claims, or to build theories, about the behavior of humans in general. While the studies may be useful starting points for such claims and theories, and may be useful for establishing experimental protocols, validation requires cross-cultural replication.

Behavioral science is not the only field that relies on narrow sampling populations. In the field of genomics, 96% of all genome-wide association studies (GWAS) have samples taken from people of European descent [11]. While the studies that have been conducted have improved understanding of genetic influence on diseases, and have highlighted genetic risk factors for many of these diseases, researchers understand that the homogeneity of the research population limits the generalizability of findings. Even after a decade-long push to diversify samples, Popejoy and Fullerton found in a 2016 study that sample-biases still favor people with European ancestry (86%) [12].

On a more limited scale, the results reported in this paper suggest that American physics education research often falls into the same trap: We have relied on research subjects who in many respects are not representative of the overall population of physics students. As a result, we need to be very careful about the assumptions that we make when interpreting the data that we have. While we are not aware of researchers explicitly claiming that their results are generalizeable to the overall population of American physics students, there is usually an implicit expectation that the results obtained by a researcher in one physics classroom should be replicable in another physics classroom – we have taken to calling this the zeroth law of PER because it is an unspoken assumption that underlies much of our discourse. Our results suggest that we as a community don't have enough data to confidently extend our research results to a majority of physics students, including high school students, students in two-year colleges, and students at less selective four-year institutions.

The idea for this paper originated in the experiences of one of the authors (SK) after he moved institutions and tried to implement PER-based instructional materials in a new setting. While the materials were unquestionably worthwhile, and led to significant and measureable improvements in student understanding, they were less effective than expected based on previous use. To what extent was this due to a difference in student preparation? While there were other differences in implementation that might also account for the observed differences in performance, it was not obvious how to go about determining the cause of the change. Because the prevalence of PER-based materials have been developed on the basis of research conducted on better-prepared students, it is difficult for an individual instructor to determine whether PER-based materials are going to be as effective with students who are less prepared. In addition, because PER has not looked in detail at variations in student populations, it is not really clear what *better preparation* for physics courses really means. Is it simply more effective traditional mathematics preparation? Are there correlations with scientific reasoning skills? Are there correlations with other cognitive skills such as spatial reasoning? Does *better preparation* primarily mean a difference in epistemological development and stance? Does it depend on past experience with more challenging problems? Are there strong cultural effects? These are all open, interesting, and probably difficult questions, and an improved understanding of the differences in student responses from one population to another will help to answer them.

How should the physics education research community move toward accounting for population differences? A helpful first step would be improving our research community's characterization of our research subjects: We should be more explicit about who our populations are, including details about them that we think influence the results we are seeing. This will allow the PER community, and others, the ability to assess the generalizability of our research findings. Currently, the description of institution and population sample is often vague, for example "Our study was conducted at a large Midwestern university," and sometimes we don't even learn what courses the students were enrolled in. If we hope to gain a better understanding of the variation in student responses from one population to another, it is important to undertake a characterization of the research population that provides a basis for comparison. This involves some guesswork about the factors that may be relevant – for example, it might be that for some studies race, gender, and sexual orientation are very important, while for other studies they are not. We expect that the physics education research community will become more adept at selecting relevant characteristics with an increase in focus on variation across populations, and as more research is conducted with a less homogeneous student population.

We suspect that part of the reluctance to offer more details about our student populations and their preparation stems from a recognition that as researchers we have an ethical obligation to protect the privacy of our research subjects, and possibly also from a desire to shield our institutions from potential embarrassment if student responses are not what we might hope for. However, the tendency to be overly vague is not without cost. A potential adopter of materials needs to be able to understand what the sample populations are in order to evaluate how the results can apply to their population of students. Without sufficient information about the research population, how can we understand the results and then assess if we would or would not expect similar results with *our* populations of students?

More detailed descriptions of our research populations are only a necessary first step. In the long run, it is important for our research community to make sure that we are not ignoring entire groups. We should ensure that we have chosen research subjects in a manner that ensures that our results apply to all students who might potentially benefit. The Belmont Report [13] that summarizes the basis for ethical research involving human subjects includes a 'principle of justice' that makes this obligation clear:

> *The choice of participants in research needs to be considered carefully to ensure that groups (e.g., welfare patients, particular racial and ethnic minorities, or persons confined to institutions) are not selected for inclusion mainly because of easy availability, compromised position, or manipulability.*
>
> *In order to achieve an equitable distribution of the risks and potential benefits of the research, investigators must determine the distribution of different groups (men and women, racial or ethnic groups, adults and children, age, etc.) in the populations that … are anticipated to benefit from the knowledge gained through the research.*

While individual Institutional Review Boards probably work to ensure that equitable choices of research subjects are made *within* each institution, the fact that PER is typically conducted at more selective and homogenous institutions means that as a nationwide systemic issue, there is little that is done to ensure equitable distribution of the benefits of what is usually publically-funded research.

Our focus on students who are wealthier, whiter, and better mathematically prepared than the overall student population reinforces the notion that students that we study are 'normal' physics students and other students are deviations from this norm. (This mindset was exemplified by Supreme Court Justice Roberts' question in *Fisher vs University of Texas* about what diverse students bring to a physics classroom – a question that contains an implicit understanding that a default physics classroom contains no racially and/or ethnically diverse students. [14]) This framing goes hand-in-hand with a 'deficit model' of racially and/or ethnically diverse student performance in physics classes, where analysis of racially and/or ethnically underrepresented students' participation is viewed in terms of how they compare to white students. With this perspective, racially and/or ethnically diverse students are likely to be framed as lacking in some physics trait when compared to the implied "norm," and intervention involves changes to instruction to 'normalize' these students. It is more useful to approach differences in results as a function of population as simply that: differences. We should not be implicitly trying to understand how to make our population more like some assumed 'norm' when differences do emerge.

The relative racial, gender, and socio-economic homogeneity of the overall physics community is an issue that in recent years is an increasing focus of the physics education research community. The causes of this lack of equity are systemic and numerous, and obviously go beyond considerations of participation in research. However, we believe that working towards equity in our research would be a great contribution to promoting inclusiveness in physics and in the disciplines that require physics, as this equity will provide the baseline information that we need to increase the likelihood that *all* students are more successful in their physics courses and are thus more likely to pursue physics-related careers.

*Conclusion*

In many ways, results of physics education research have shaped the way education has evolved over the past few decades -- from content, to delivery, to classroom layouts, etc. However, as a research community, we have not been sufficiently attentive to whether these robust, impactful results have applied to *all* students. We have implicitly assumed that the populations that we have researched are representative. But the preparation, motivations, and goals of students at a non-research university in a predominantly rural state are likely quite different from those of students raised in cities and attending highly-competitive research institutions. Moreover, the resources available to instructors at these institutions are likely to be different as well. Published studies that gloss over the disparities that exist between these groups of students might compare the performance of these groups of students without any discussion of issues related to population. This leads to results that are likely not reproducible and that do not form a solid basis for future research.

We do not argue that the results obtained and the theoretical models proposed by physics education researchers lack value because they have failed to take population differences into account and because they have tended to focus on a well-prepared homogeneous population. A useful analogy can be made to the sequence of research in a given field, where initial studies are often restricted to situations where phenomena are simplified and many variables are eliminated. For example, early climate change studies ignored effects of interactions between oceans and the atmosphere, and did not include effects of changes to the planet's biomass [15]. Only after the models developed with these simplifying assumptions were better understood did researchers start to include more complicated and realistic scenarios. Similar strategies are employed when we teach physics: When Newton's Laws are taught, for example, first we teach students in idealized situations without resistive effects. Once students have learned how to apply Newton's laws in simple situations, we add a few more selected variables such as friction and air resistance, and then we might include multiple dimensions or different coordinate systems.

Physics education research started with what we consider to be a simplified situation that included well-prepared calculus-based students with relatively homogenous and privileged backgrounds. From our perspective, the physics education research community is now at a point in the overall research enterprise where we would greatly benefit by explicitly including population variability in our studies. It is possible that our results remain unchanged when many of the studies that we have already conducted are repeated with different student populations. On the other hand, we may see dramatically different results, which will allow us to focus on the causes of these differences. We expect that additional influences to overall findings will be discovered, much in the same way that adding friction to a mechanics problem changes final velocities without eliminating the known effects of mass and gravitation. We hope that the PER community increasingly approaches studies with the question "What differences in results are due to population differences?"

*Acknowledgements*

The authors would like to thank Susan White and Patrick Mulvey and the American Institute of Physics Statistical Research Center for their help with compiling statistics for overall physics enrollments.